\documentclass[twocolumn,showpacs,amsmath,amssymb]{revtex4}

\usepackage{graphicx}
\usepackage{bm}

\def\D{\partial}
\def\grad{\nabla}

\def\lap{\nabla^2}
\def\inv{^{-1}}

\def\Av.#1{\overline{#1}}
\def\Eq#1{(\ref{eq:#1})}
\def\eql#1{\label{eq:#1}}
\def\Fig#1{Fig.\ref{fig:#1}}   
\def\figl#1{\label{fig:#1}}

\def\eq{\begin{eqnarray}}
\def\qe{\end{eqnarray}}

\def\eqnn{\begin{eqnarray*}}
\def\qenn{\end{eqnarray*}}

\def\nn{\nonumber}

\def\be{\bm{e}}

\def\bn{\bm{n}}

\def\br{\bm{r}}

\def\simge{\;\lower3pt\hbox{$\stackrel{\textstyle >}{\sim}$}\;}
\def\simle{\;\lower3pt\hbox{$\stackrel{\textstyle <}{\sim}$}\;}
\def\hsp#1{\hspace{#1mm}}
\def\vsp#1{\vspace{#1mm}}

\def\->{\rightarrow}
\def\=>{\Rightarrow}
\def\<->{\leftrightarrow}
\def\<<{\ll}
\def\>>{\gg}

\def\bm#1{\mbox{\boldmath $#1$}}

\def\lrL#1{\left[#1\right]}

\def\lrS#1{\left(#1\right)}
\def\lrF#1{\left|#1\right|}

\def\mycomment#1{}

\def\f#1#2{\frac{#1}{#2}}

\def\lam{\lambda}
\def\Fsm{F_{\rm{sm}}}
\def\Fcp{F_{\rm{cp}}}
\def\Ffr{F_{\rm{Fr}}}
\def\p0{\psi_0}
\def\pb{\bar \psi}
\def\bxi{\bar \xi}
\def\fTGBA{f_{\rm TGBA}}
\def\fMGBA{f_{\rm MGBA}}
\def\kcl{k_{\rm c1}}
\def\kdcl{k^{\rm RL}_{\rm c1}}

\def\ac{\alpha_c}

\begin{document}

\preprint{APS/123-QED}

\title{Selection of defect structures in twist-grain-boundary-A phase of chiral liquid crystals}

\author{Hiroto Ogawa}
\affiliation{%
Department of Physics, Tohoku University, Sendai, 980-8578, Japan
}%
\email{hiroto@cmpt.phys.tohoku.ac.jp}


\date{\today}

\begin{abstract}
We study the structure of the twist grain boundary in chiral smectic liquid crystals using the Landau-de Gennes model. By
considering spatial variation of the smectic order, we distinguish the Melted-Grain-Boundary-A (MGBA) structure with cholesteric-like
domains and the Twist-Grain-Boundary-A (TGBA) structure consisting of screw dislocations. The MGBA structure becomes
stable near the transition
to the cholesteric phase. On approaching the transition,
the cholesteric-like domain grows outside the grain boundary.
Also, the dislocation spacings of the TGBA structure agree better with experimental results than previous theories.
\end{abstract}

\pacs{Valid PACS appear here}
\maketitle

\par
Nontrivial physics in condensed matter is often caused by resolution of frustration.
For example, mismatch between the triangular crystalline order and antiferromagnetic spin order
disturbs an equilibrium ordered phase even at the absolute zero~\cite{Spin}.
In the glass system comprising two types of hard core particles with different radii,
the positional order of the particles
and the bond order defined by the number of the nearest neighbor particles
are frustrated, and generate disclinations~\cite{Rubinstein}.

Liquid crystals exhibit a wide variety of frustration-induced defect structures~\cite{deGennes, Lubensky}.
One of the reasons is the various kinds of order they show: nematic order, smectic layering order,
and cholesteric (helical) order due to molecular chirality.
Among them, frustration between the smectic and cholesteric order is paid great attention
because of scientific interest and potential application.
Twist-grain-boundary (TGB) phase is the simplest one-dimensional defect phase,
comprising series of finite-length smectic slabs (grains)
with the layer normals rotating at a constant twist angle $\alpha$,
intervened by grain boundaries (\Fig{TMGB}(a)).
It was predicted in the pioneering work~\cite{Renn}
in terms of the analogy between liquid crystals and superconductors,
and confirmed by experiment~\cite{Goodby}.
As temperature increases, the ordered lattice of the screw dislocations "melts" into the dislocation fluid,
called the chiral line (NL$^*$) phase~\cite{NL}.
At higher temperature, defects construct a variety of three-dimensional networks,
called smectic blue phases~\cite{Kitzerow}.
Although the phase transitions between these defect phases are observed,
their spatial structures are not yet well understood.
\par
The grain boundary in the TGBA phase,
having the molecular orientation parallel to the layer normal in the smectic slab,
has been thought to consist of parallel screw dislocations,
in analogy to the magnetic fluxes in the vortex lattice phase of superconductors.
This structure (denoted by the TGBA structure) was confirmed with the transmitted electron microscopy (TEM) observation~\cite{Ihn}.
The length of the smectic slabs $l_b$ and the interval of the screw dislocation $l_d$
were experimentally estimated~\cite{Presumed},
and theoretically calculated~\cite{Linear, SantangeloPRE}.
However, the theoretical values are about 20 times higher than the experimental values.
This discrepancy in the defect
spacings might be
because of the theoretical treatment of the defect energy.
\par
Recently, another possible structure of the grain boundary has been proposed~\cite{PNAS}.
In the TEM study, the screw dislocation array is not observed,
which implies that
the smectic order (and hence the dislocation array) is melted in the grain boundary (\Fig{TMGB}(b)).
This structure is
called
the melted-grain-boundary-A (MGBA) structure,
and is distinguished from
the TGBA which has the periodic structure in the grain boundary.
\par
In this Letter, we consider the spatial variation of the smectic order,
to analyze the structure of and compare the free energy of
the TGBA and MGBA states.
As by-products, we will calculate the characteristic lengths $l_b$ and $l_d$ in the TGBA state,
and fix the discrepancy between theory and experiment.
\begin{figure}[t]
\begin{center}
\begin{tabular}{ll}
(a)&
(b)\\
\includegraphics[width=0.2\textwidth]{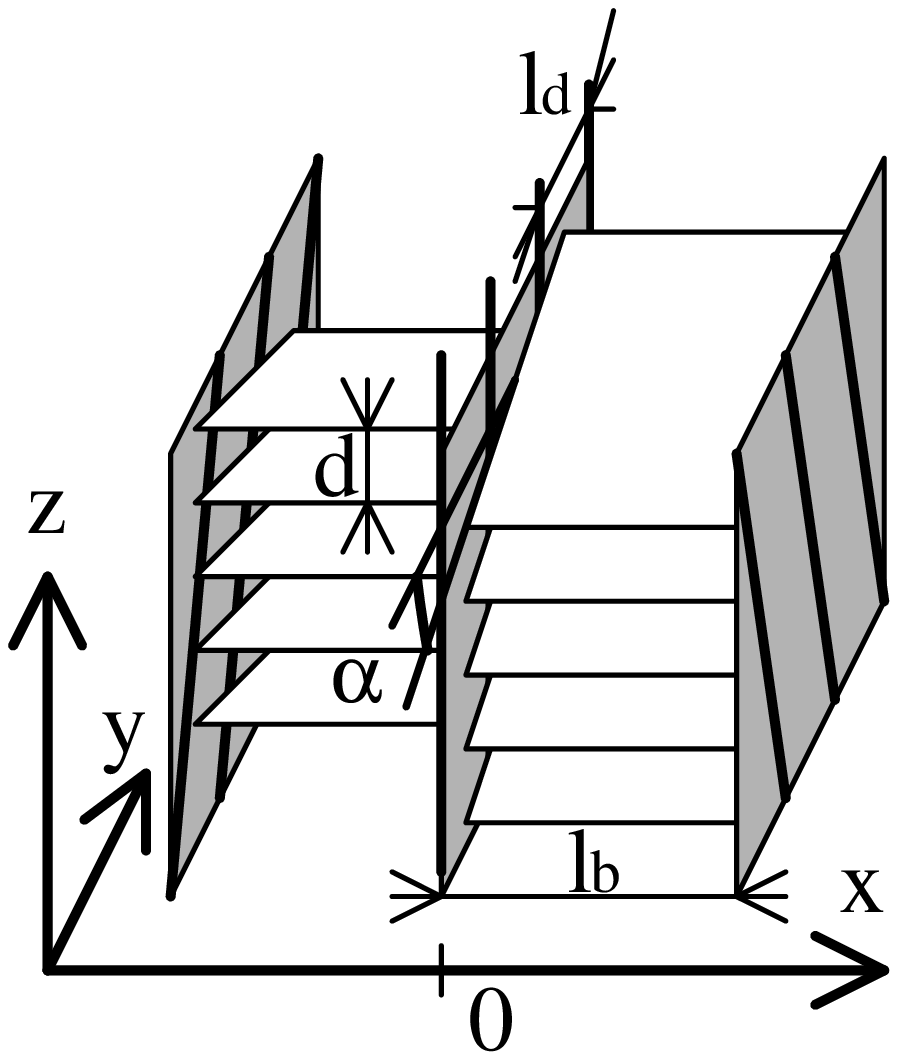}&
\includegraphics[width=0.2\textwidth]{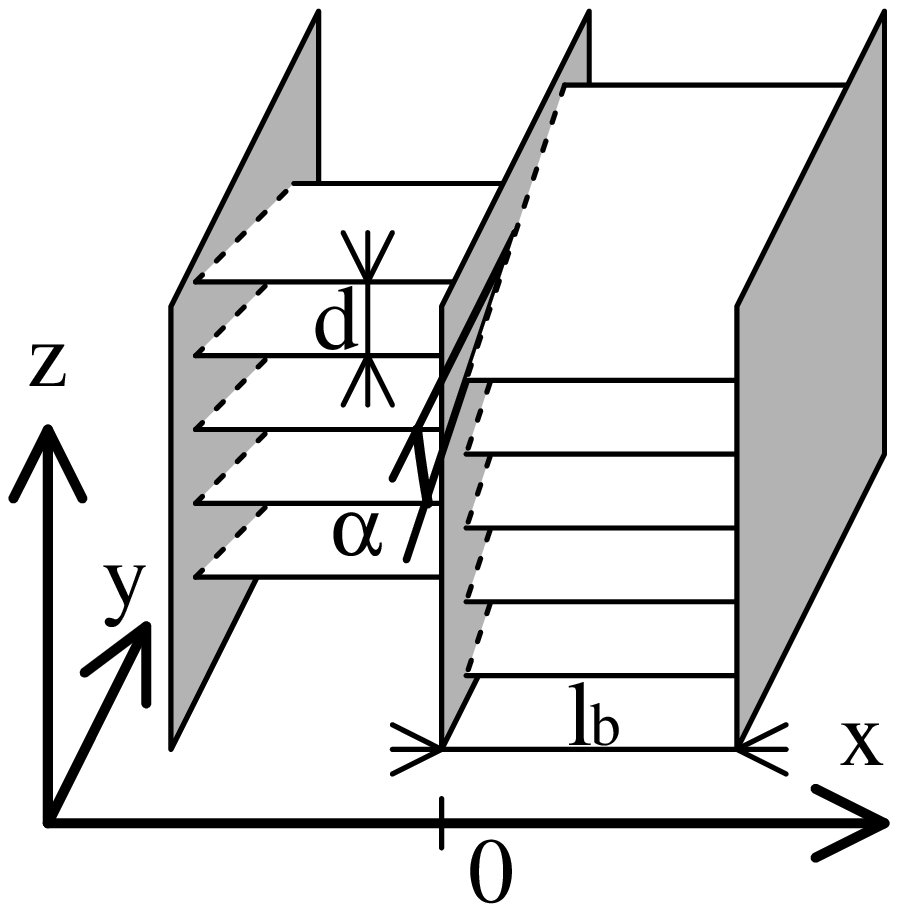}
\end{tabular}
\end{center}
\caption{Schematic representations of the TGBA phases.
(a) The TGBA structure consists of the parallel screw dislocations (bold lines).
(b) The MGBA structure contains no screw dislocation.}
\figl{TMGB}
\end{figure}
\par
We consider one pair of the adjacent grain boundaries,
and the
smectic slab between them (\Fig{TMGB}).
Setting the $x$-axis along the twist axis,
we assume that the director $\bn$ lies only in the $yz$-plane~\cite{Renn},
and
express it by the local twist angle $\omega$ measured form 
the middle of the two grain boundaries
as $\bn=\lrS{0, -\sin\lrS{\omega+\alpha/2}, \cos\lrS{\omega+\alpha/2}}$.
The smectic order is expressed by a complex order parameter $\Psi$,
and the equilibrium wave number of the smectic density wave is denoted by $q_0$.
Layer displacement $u$ is represented as $\Psi=\psi\exp\lrL{iq_0\lrS{z-u}}$,
where the z-axis is set along the layer normal at $x=0$,
and $\psi$ is the strength of the smectic order.

The equilibrium structure is obtained by
minimization of the Landau-de Gennes free energy
consisting of the three components:
\eq
F_{\rm{sm}}&=&\int d\br\f{1}{2\xi^2\p0^2}\lrS{|\Psi|^2-\psi_0^2}^2,
\eql{Fsm}\\
F_{\rm{Fr}}&=&\int d\br\lam^2q_0^2\psi_0^2\lrS{\D_x\omega-k_0}^2.
\eql{Ffr}\\
F_{\rm{cp}}&=&\int d\br\lrF{\lrS{\grad-iq_0\bn}\Psi}^2,
\eql{Fcp}
\qe
The smectic energy $\Fsm$ adjusts $\lrF{\Psi}$
to the equilibrium value $\p0$.
$\Ffr$ is the Frank elastic energy of $\bn$,
where the three elastic moduli of the splay, twist, and bend components
are set to be equal.
$k_0$ is the inversed helical pitch due to the molecular chirality.
The coupling energy $\Fcp$ is the cross term of $\Psi$ and $\bn$.
The correlation lengths of $\Psi$ ($\bn$) is $\xi$ ($\lam$),
proportional to $1/\sqrt{T_{\rm NA}-T}$. $T_{\rm NA}$ is the nematic (N)-smectic A (SmA) transition temperature
for the achiral case $k_0=0$.
 
\begin{figure}[t]
\begin{center}
\includegraphics[width=0.25\textwidth]{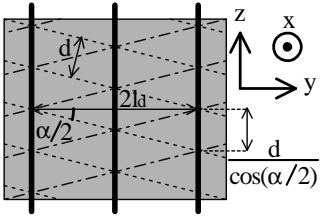}
\end{center}
\caption{The layer geometry of the TGBA structure seen from the twist axis (x-axis).
The screw dislocations (bold lines) are supported by the smectic layers
in front of the grain boundary (dot-dash lines) and behind the grain boundary (dashed lines).}
\figl{GB}
\end{figure}

For the TGBA,
the smectic order is destructed near the dislocations
by the helical order.
In the grain boundary with the thickness $\sim\xi$,
the director changes slowly so that $\bn\simeq\be_z$ (\Fig{GB}),
because $\xi\ll\lam$ in typical experiments~\cite{PNAS}.
The layer displacement is approximated with the linear superposition of the screw dislocations~\cite{Kamien}
\eq
uq_0\cos\f{\alpha}{2}&=&-\sum_{n=-\infty}^{\infty}\tan\inv\f{y-nl_d}{x}\nn\\
&=&-\f{1}{2}{\rm Im}\ln\sinh\f{\pi\lrS{x+iy}}{l_d}.
\eql{layer_tgb}
\qe
This layer distortion characterize the screw dislocation array,
and $\Fcp$, containing $u$, is dominant near the grain boundary.
Close to the dislocation cores,
the local profile of $\psi$ is a solution of $\delta \Fcp/\delta \psi=0$,
which is approximated as
\eq
\lap\psi=\lrS{\f{\pi}{l_d}}^2\f{\cosh\lrS{2\pi x/l_d}+\cos\lrS{2\pi y/l_d}}
{\cosh\lrS{2\pi x/l_d}-\cos\lrS{2\pi y/l_d}}\psi.
\eql{eq_tgbp}
\qe
The solution is
\eq
\psi=C\sqrt{\cosh\lrS{2\pi x/l_d}-\cos\lrS{2\pi y/l_d}},
\eql{tgbp}
\qe
where $C$ is an integration constant~\cite{OgawaPhD}.
Near the SmA-TGBA transition,
$\psi$ increases up to $\p0$ outside the grain boundary $x>\xi$.
The energy density cost
of making dislocation is
$\epsilon=2\pi\p0^2l_d\alpha/2$,
and the energy density gain by twisting the layers is $\p0^2q_0^2\lam^2k_0\alpha$.
Equating the two contributions,
we determine the lower critical field for the SmA-TGBA phase transition $k_0=\kcl=1/2q_0\lam^2$,
which is much lower than the previous Renn-Lubensky result $\kdcl=\ln\lrS{\lam/\xi}/2q_0\lam^2$~\cite{Renn}.
In our theory, the energy per
dislocation is lowered,
because the long range layer displacement around the single dislocation $\lrS{|\grad u|\propto 1/r}$
is
summed out
\Eq{layer_tgb},
and also because of the relaxation of $\psi$
whose correlation length $\xi$ is shorter than that of the director $\lam$,
which has been neglected in previous theories.

For the MGBA structures,
$\psi$ is uniformly small in the
grain boundary.
The solution of the equilibrium condition $\delta \Fcp/\delta \psi=0$
is
\eq
\psi=D\sin\f{x}{\xi},
\eql{mgbp}
\qe
where $D$ is an integration constant~\cite{OgawaPhD}.

The energy of the grain boundary thus can be calculated with \Eq{tgbp} for the TGBA,
and \Eq{mgbp} for the MGBA.
It can be shown that the director rotates at a constant rate because of the weak smectic order.

For both the TGBA and MGBA, the smectic order grows
and $\Fsm$ becomes one of the main contributions
at distance larger than $\xi$ from the grain boundary.
In the smectic slab $x>\xi$,
the twist is expelled,
and the layer normal $\be_z+\grad u$ is close to a constant $\lrS{0, -\sin\alpha/2, \cos\alpha/2}$.
Using this relation for $u$,
the sum of the smectic and coupling energies is reduced to the effective smectic energy
\eq
\Fsm+\Fcp&=&\int d\br\lrL{\lrS{\grad\psi}^2+\f{1}{2\xi^2}\lrS{\psi^2-\pb^2}^2},
\eql{Fsmcp}\\
\pb&=&\p0\sqrt{1-4q_0^2\xi^2\sin^2\f{\omega}{2}},
\eql{pb}
\qe
where $\pb$ is the effective equilibrium smectic order,
and $\omega$ ranges from $-\alpha/2$ to $\alpha/2$.
Because of the frustration between the smectic and helical order,
$\psi$ is reduced at large $\omega$.
According to \Eq{pb},
if $\omega$
increases beyond $\alpha_c/2=2\arcsin\lrS{1/2q_0\xi}$,
the equilibrium smectic order vanishes,
and a cholesteric-like domain appears near the grain boundary.
In the cholesteric domain, the director rotates at a constant rate, which we denote $\omega'_0$.

If the twist angle $\alpha$ is smaller than $\alpha_c$,
$\psi$ approximately equals $\pb=\pb(\omega)$ in the smectic slab $x>\xi$,
because the characteristic length scale of $\omega$ and $\pb$ is $\lam$
$(\gg \xi)$.
The twist of the director penetrates the smectic slab with the characteristic length $\lam$
in the form $\omega\propto\sinh\lrL{\lrS{x-l_b/2}/\lam}$.
Combining the solutions for the grain boundary and the smectic slab,
we obtain the total free energy densities
for the TGBA
and MGBA
states, as
%
%
%
\eq
&&\hsp{-29}
\fTGBA/\p0^2=\f{1-P_T^2(\kappa\tanh X+1)}{2\xi^2(\kappa X+1)},
\eql{fTGBA}
\qe
and
\eq
&&\hsp{-8}
\fMGBA/\p0^2=
\f{1}{2\xi^2(\kappa X+1)}\nn\\
&&\hsp{-5}
\times\lrL{1+2\cot1-\f{P_M^2}{\lrS{1+\kappa\tanh X}\inv-\kappa^{-2}\cot1}},
\eql{fMGBA}
\qe
where the dimensionless chiralities $P_T=\lrS{k-\kcl}/k_c$ and $P_M=k/k_c$ are introduced,
and $X=\lrS{-\xi+l_b/2}/\lam$ is the ratio between the length of the smectic domain
and the director correlation length.
\par
Minimization of the free energy density \Eq{fTGBA} determines $l_b$ and $l_d=d/2\sin\lrS{\alpha/2}$
for the TGBA state
as a function of the thermodynamic variables $T$ and $k_0$,
To the lowest order of $\kappa\inv$,
the equilibrium structural lengths are obtained as $l_b/\lam=2\lrS{X+\kappa\inv}$,
and $l_d/\lam=\sqrt{2}\pi/P_T\lrS{1+\kappa\tanh X}$,
where $\kappa X=\sqrt[3]{3/\beta}\lrL{\sqrt[3]{1+\sqrt{1-\beta/3}}
+\sqrt[3]{1-\sqrt{1-\beta/3}}}\inv$
and $\beta=\lrS{P_T/\kappa}^2$.
In typical experiments~\cite{Presumed, Ihn}, $P_T\sim 0.1$, and $\kappa\sim 100$.
The ratio $l_b/l_d$ remains on the order of 1,\
agreeing with both the experimental and previous theoretical results~\cite{Linear, SantangeloPRE, Presumed}.
The order of $l_b/\lam$ and $l_d/\lam$ is $0.1-0.01$ except near the SmA-TGBA phase transition,
which agrees again with the experimental results~\cite{Presumed}.
In our theory, introducing the spatial variation of $\psi$,
the defect energy is evaluated much lower
(so that the equilibrium defect density is larger) than in the previous theories~\cite{Linear, SantangeloPRE}.

Change of the stability between the TGBA and MGBA states occurs when $\fTGBA=\fMGBA$.
This condition is satisfied for $P_T/\kappa, P_M/\kappa\simeq 0.10$, which corresponds to the twist angles
$\alpha\simeq 0.41\ac$ for the TGBA and $\alpha\simeq 0.56\ac$ for the MGBA.
Averaging the conditions, in experiment, a TGBA-like structure would change to a MGBA-like structure at $\alpha=\alpha_{\rm TM}\simeq 0.5\ac$,
or $l_d\simeq 6\xi$.
This means that the MGBA can be understood
as overlapped screw dislocations each of which has core radius on the order of $\xi$.
In a real situation, transition between the TGBA and MGBA might not be thermodynamic.
and
could be a crossover with a gradual change of the grain boundary structure.
Note that here we compared the two limiting cases (the TGBA with the linearly superposed dislocations
and the MGBA with the perfectly uniform grain boundary).

If the twist angle is large such that $\alpha>\alpha_c$,
a local cholesteric-like domain appears where $|\omega|>\alpha_c/2$.
Existence of the cholesteric domain means that the smectic order melts at the whole grain boundary.
Thus if $\alpha>\alpha_c$, the structure is {\it always} MGBA.
We obtain an approximated equation for $\psi$ in the cholesteric domain as
$\psi''=q_0^2\psi\lrS{\omega_0'x-\alpha/2}^2-\psi/\xi^2$,
with the help of $\sin\lrS{\omega/2}\simeq\omega/2$ for $|\omega|/2<\pi/4$,
and $q_0^2\xi^2\gg 1$ for the typical case~\cite{PNAS}.
In the London limit~\cite{Renn, Linear}, using $\lam\gg\xi\gg q_0\inv$,
the solution is approximated as
$\psi\propto|\eta|^{\lrL{\lrS{\omega_0'q_0\xi^2}\inv-1}/2}\exp\lrS{-\eta^2/2}$,
where $\eta$ is the reduced coordinate $\lrS{x-\alpha/2\omega'_0}/\bxi$
with the effective correlation length $\bxi=1/\sqrt{q_0\omega'_0}$.
Defining $L$ by $\omega\lrS{x=L}=-\alpha_c/2$,
the cholesteric domain extends to $x=L+\xi$,
because $\pb\lrS{L+\xi+\delta x}\sim\p0\lrS{q_0\xi}^{1/2}\lrL{\lrS{\xi+\delta x}/\lam}^{1/2}$
grows
to the order of $\p0$ as $\delta x$ increases to the order of $\lam$.
On the other hand, the smectic domain satisfying $\psi=\pb$ ends at $x=L+\xi$,
because $\lrL{\pb'\lrS{L+\xi}/\p0}\inv\sim\xi$ and the gradient term in \Eq{Fsmcp}
becomes a major contribution.
Combining the three domains, the grain boundary, the cholesteric domain, and the smectic domain,
we obtain the total free energy density.
The smectic domain is vanished at the transition to the cholesteric (N*) phase,
and the upper critical field $k_{c2}$ is shown to have the same value $1/q_0\xi^2$
as in the Renn-Lubensky result~\cite{Renn}.
The length of the grain boundary and cholesteric domain is plotted in \Fig{chol}(a).
The cholesteric domain starts to grow more rapidly than $\lam$
at the temperature where the twist angle $\alpha$ equals the critical angle $\alpha_c$.
Thus if the twist angle for this domain growth is measured,
one can indirectly estimates the characteristic angle of change from the TGBA to MGBA $\alpha_{\rm TM}\simeq 0.5\alpha_c$.
Note, the abrupt growth of the cholesteric domain in \Fig{chol}(a) might be an artifact.
Instead, the growth should be continuous~\cite{OgawaTGB}.

The phase diagram is shown in \Fig{pd}(b).
The equilibrium state changes as SmA-TGBA-MGBA-N*
with the increment of the temperature and chirality.
This state sequence can be understood in terms of symmetry.
Both of the SmA and TGBA have the discrete translational symmetry of the smectic order in the $yz$-plane.
Only the pair of the TGBA and MGBA possesses the discrete helical symmetry along the x-axis.
For the MGBA and N*, they are only the pair having the continuous symmetry in the $yz$-plane.
The state pairs having the common symmetry are neighboring in the phase diagram.
In addition, stability of the TGBA state to the SmA phase is higher than that in the previous work,
because of the energy cost of the grain boundary is
lowered by
the interaction between
dislocations in
each grain boundary,
as well as by the relaxation of
the smectic order parameter.
Such changes of the defect energy and structure by introducing the spatial variation of the smectic order
may occur also in the TGB-embedded double twist cylinder~\cite{KamienDT} and other smectic blue phases~\cite{Kitzerow}.
The TGB phase has a simple one-dimensional defect structure,
and our results may provide a fundamental clue
to understand the structures of these complex defect phases.
%
We also note that the MGBC* phase was shown to be always stable over the TGBC* phase
due to the macroscopic helical order of the director along the layer normal~\cite{Dozov}.
The MGBA structures does not have such a helical order and its stability over the TGBA
was far from trivial.
%


\begin{figure}[tphb]
\begin{tabular}{ll}
(a)&(b)\\
\begin{minipage}{0.19\textwidth}
\vsp{4.5}
\includegraphics[width=\textwidth]{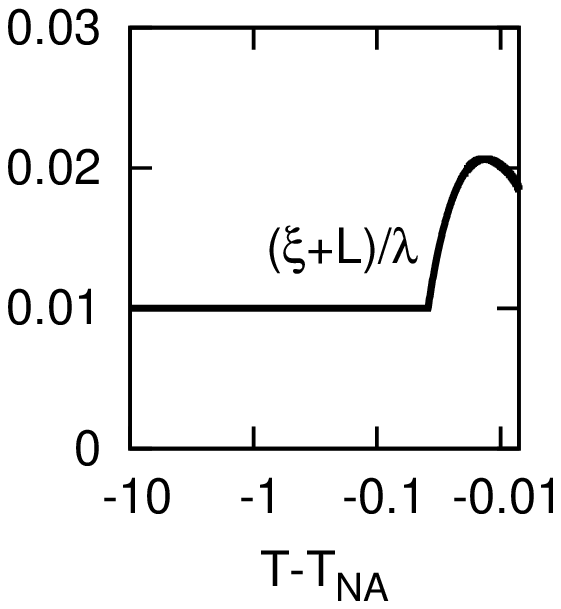}
\end{minipage}
&
\begin{minipage}{0.20\textwidth}
\includegraphics[width=\textwidth]{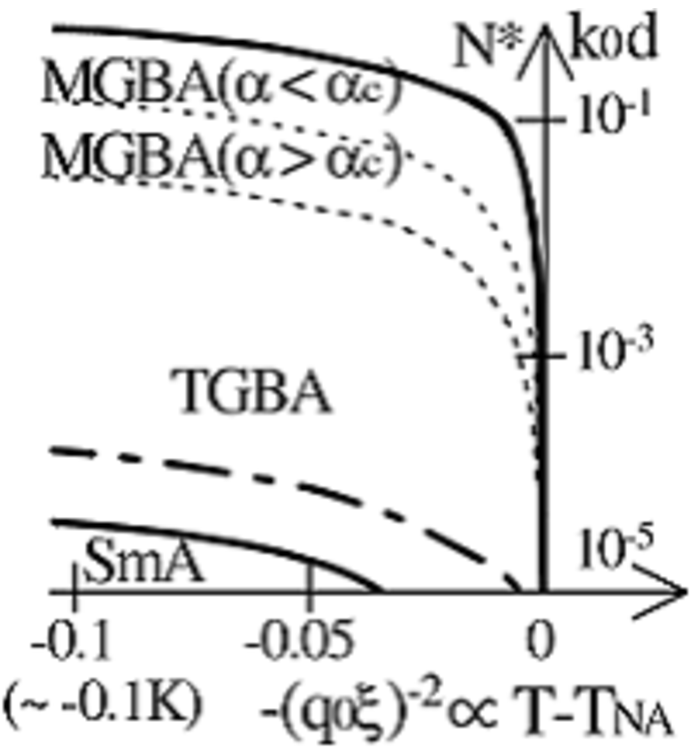}
\end{minipage}
\end{tabular}
\caption{(a) The temperature dependence of the size
of the grain boundary and cholesteric domain (divided by $\lam$).
The temperature range between the SmA-TGBA and MGBA-N* transition is shown.
(b)The phase diagram. Solid lines mean the second order phase transition.
The dot-dash line shows the previous result for the SmA-TGBA transition~\cite{Renn}.
Dashed lines separate different equilibrium structures.
}
\figl{chol}
\figl{pd}
\end{figure}

Let us discuss our results in terms of the analogy between
the TGB phase and
the vortex lattice phase of superconductors.
The smectic order parameter $\Psi$ and the director $\bn$
correspond to the wave function of the superconducting particles and the vector potential, respectively.
The screw dislocation in the TGB phase, the twist angle $\alpha$, and the molecular chirality $k_0$
correspond to the vortex, the magnetic flux, and the external magnetic field.
The difference is the dimensionality:
In the TGB phase, three spatial dimensions are
divided into one for the direction of the screw dislocation,
one for the twist axis, and one for the periodic dislocation array.
In the vortex lattice phase of isotropic superconductors, on the other hand,
the three are divided
into one for the direction of the vortex, and two for the lattice periodicity.
However, in anisotropic superconductors,
vortices are aligned in a one dimensional chain for a certain magnetic field range as in the TGB phase.
In liquid crystal, the origin of the corresponding anisotropy is the molecular chirality.
Theoretical study of the vortex chain is based on the interaction of the phase of the wave function,
corresponding to the layer displacement in liquid crystals~\cite{Koshelev}.
Thus it is interesting to introduce the spatial variation of wave function near the vortices,
as in our theory for chiral liquid crystals.
The vortices in a chain are "melted" to the Josephson vortex by stronger anisotropy.
Thus one might regard the MGBA state as an analogue to the Josephson vortex state.

In summary, we analyze the TGBA and MGBA states of chiral liquid crystals with the Landau-de Gennes model.
By considering the spatial variation of the smectic order parameter,
we found that a cholesteric-like domain appears at twist angle higher than the critical angle $\alpha_c$.
We have shown that the proposed MGBA state~\cite{PNAS}
is certainly stable over the TGBA state
when
the twist angle is higher than $\alpha_{\rm TM}\simeq 0.5\alpha_c$.
We may
indirectly specify the MGBA state
through observing
the cholesteric domain
that appears
at $\alpha=\alpha_c$.
The energy cost of
the dislocation is also greatly changed from the previous theory,
resulting in an agreement of the dislocation spacings $l_b$, $l_d$ with the experiments~\cite{Presumed},
and the enhanced stability of the TGBA state over the SmA phase.
It would be interesting to extend our results
to other defect phases, including the smectic blue phases of liquid crystals,
and the vortex chains of anisotropic superconductors.

The author thanks Nariya Uchida,
Toshihiro Kawakatsu, Jun Yamamoto, Katsuhiko Sato, and Yoshinori Hayakawa for fruitful discussions.
This study is supported by JSPS fellowship.

\bibliography{thesis}

\begin{thebibliography}{19}
\expandafter\ifx\csname natexlab\endcsname\relax\def\natexlab#1{#1}\fi
\expandafter\ifx\csname bibnamefont\endcsname\relax
  \def\bibnamefont#1{#1}\fi
\expandafter\ifx\csname bibfnamefont\endcsname\relax
  \def\bibfnamefont#1{#1}\fi
\expandafter\ifx\csname citenamefont\endcsname\relax
  \def\citenamefont#1{#1}\fi
\expandafter\ifx\csname url\endcsname\relax
  \def\url#1{\texttt{#1}}\fi
\expandafter\ifx\csname urlprefix\endcsname\relax\def\urlprefix{URL }\fi
\providecommand{\bibinfo}[2]{#2}
\providecommand{\eprint}[2][]{\url{#2}}

\bibitem[{\citenamefont{Vannimenous and Toulouse}(1977)}]{Spin}
\bibinfo{author}{\bibfnamefont{J.}~\bibnamefont{Vannimenous}} \bibnamefont{and}
  \bibinfo{author}{\bibfnamefont{G.}~\bibnamefont{Toulouse}},
  \bibinfo{journal}{J. Phys. C} \textbf{\bibinfo{volume}{10}},
  \bibinfo{pages}{537} (\bibinfo{year}{1977}).

\bibitem[{\citenamefont{Nelson et~al.}(1982)\citenamefont{Nelson, Rubinstein,
  and Spaepen}}]{Rubinstein}
\bibinfo{author}{\bibfnamefont{D.~R.} \bibnamefont{Nelson}},
  \bibinfo{author}{\bibfnamefont{M.}~\bibnamefont{Rubinstein}},
  \bibnamefont{and} \bibinfo{author}{\bibfnamefont{F.}~\bibnamefont{Spaepen}},
  \bibinfo{journal}{Phil. Mag. A} \textbf{\bibinfo{volume}{46}},
  \bibinfo{pages}{105} (\bibinfo{year}{1982}).

\bibitem[{\citenamefont{de~Gennes and Prost}(1994)}]{deGennes}
\bibinfo{author}{\bibfnamefont{P.~G.} \bibnamefont{de~Gennes}}
  \bibnamefont{and} \bibinfo{author}{\bibfnamefont{J.}~\bibnamefont{Prost}},
  \emph{\bibinfo{title}{The Physics of Liquid Crystals}}
  (\bibinfo{publisher}{Clarendon, Oxford}, \bibinfo{year}{1994}).

\bibitem[{\citenamefont{Chaikin and Lubensky}(1995)}]{Lubensky}
\bibinfo{author}{\bibfnamefont{P.~M.} \bibnamefont{Chaikin}} \bibnamefont{and}
  \bibinfo{author}{\bibfnamefont{T.~C.} \bibnamefont{Lubensky}},
  \emph{\bibinfo{title}{Principles of Condensed Matter Physics}}
  (\bibinfo{publisher}{Cambridge University Press, Cambridge, England},
  \bibinfo{year}{1995}).

\bibitem[{\citenamefont{Renn and Lubensky}(1988)}]{Renn}
\bibinfo{author}{\bibfnamefont{S.~R.} \bibnamefont{Renn}} \bibnamefont{and}
  \bibinfo{author}{\bibfnamefont{T.~C.} \bibnamefont{Lubensky}},
  \bibinfo{journal}{Phys.\ Rev.\ A} \textbf{\bibinfo{volume}{38}},
  \bibinfo{pages}{2132} (\bibinfo{year}{1988}).

\bibitem[{\citenamefont{Goodby et~al.}(1989)\citenamefont{Goodby, Waugh, Stein,
  Chin, Pindak, and Patel}}]{Goodby}
\bibinfo{author}{\bibfnamefont{J.~W.} \bibnamefont{Goodby}},
  \bibinfo{author}{\bibfnamefont{M.~A.} \bibnamefont{Waugh}},
  \bibinfo{author}{\bibfnamefont{S.~M.} \bibnamefont{Stein}},
  \bibinfo{author}{\bibfnamefont{E.}~\bibnamefont{Chin}},
  \bibinfo{author}{\bibfnamefont{R.}~\bibnamefont{Pindak}}, \bibnamefont{and}
  \bibinfo{author}{\bibfnamefont{J.~S.} \bibnamefont{Patel}},
  \bibinfo{journal}{Nature} \textbf{\bibinfo{volume}{337}},
  \bibinfo{pages}{449} (\bibinfo{year}{1989}).

\bibitem[{\citenamefont{Kamien and Lubensky}(1993)}]{NL}
\bibinfo{author}{\bibfnamefont{R.~D.} \bibnamefont{Kamien}} \bibnamefont{and}
  \bibinfo{author}{\bibfnamefont{T.~C.} \bibnamefont{Lubensky}},
  \bibinfo{journal}{J.\ Phys.\ I\ (France)} \textbf{\bibinfo{volume}{3}},
  \bibinfo{pages}{2131} (\bibinfo{year}{1993}).

\bibitem[{\citenamefont{Kitzerow and Bahr}(2002)}]{Kitzerow}
\bibinfo{editor}{\bibfnamefont{H.~S.} \bibnamefont{Kitzerow}} \bibnamefont{and}
  \bibinfo{editor}{\bibfnamefont{C.}~\bibnamefont{Bahr}}, eds.,
  \emph{\bibinfo{title}{Chirality in Liquid Crystals}}
  (\bibinfo{publisher}{Springer-Verlag, New York}, \bibinfo{year}{2002}).

\bibitem[{\citenamefont{Ihn et~al.}(1992)\citenamefont{Ihn, Zasadzinski,
  Pindak, Slaney, and Goodby}}]{Ihn}
\bibinfo{author}{\bibfnamefont{K.~J.} \bibnamefont{Ihn}},
  \bibinfo{author}{\bibfnamefont{J.~A.} \bibnamefont{Zasadzinski}},
  \bibinfo{author}{\bibfnamefont{R.}~\bibnamefont{Pindak}},
  \bibinfo{author}{\bibfnamefont{A.~J.} \bibnamefont{Slaney}},
  \bibnamefont{and} \bibinfo{author}{\bibfnamefont{J.~W.}
  \bibnamefont{Goodby}}, \bibinfo{journal}{Science}
  \textbf{\bibinfo{volume}{258}}, \bibinfo{pages}{275} (\bibinfo{year}{1992}).

\bibitem[{\citenamefont{Navailles et~al.}(1998)\citenamefont{Navailles, Pansu,
  Gorre-Talini, and Nguyen}}]{Presumed}
\bibinfo{author}{\bibfnamefont{L.}~\bibnamefont{Navailles}},
  \bibinfo{author}{\bibfnamefont{B.}~\bibnamefont{Pansu}},
  \bibinfo{author}{\bibfnamefont{L.}~\bibnamefont{Gorre-Talini}},
  \bibnamefont{and} \bibinfo{author}{\bibfnamefont{H.~T.}
  \bibnamefont{Nguyen}}, \bibinfo{journal}{Phys.\ Rev.\ Lett.}
  \textbf{\bibinfo{volume}{81}}, \bibinfo{pages}{4168} (\bibinfo{year}{1998}).

\bibitem[{\citenamefont{Bluestein et~al.}(2001)\citenamefont{Bluestein, Kamien,
  and Lubensky}}]{Linear}
\bibinfo{author}{\bibfnamefont{I.}~\bibnamefont{Bluestein}},
  \bibinfo{author}{\bibfnamefont{R.~D.} \bibnamefont{Kamien}},
  \bibnamefont{and} \bibinfo{author}{\bibfnamefont{T.~C.}
  \bibnamefont{Lubensky}}, \bibinfo{journal}{Phys.\ Rev.\ E}
  \textbf{\bibinfo{volume}{63}}, \bibinfo{pages}{061702}
  (\bibinfo{year}{2001}).

\bibitem[{\citenamefont{Santangelo and Kamien}(2007)}]{SantangeloPRE}
\bibinfo{author}{\bibfnamefont{C.~D.} \bibnamefont{Santangelo}}
  \bibnamefont{and} \bibinfo{author}{\bibfnamefont{R.~D.}
  \bibnamefont{Kamien}}, \bibinfo{journal}{Phys.\ Rev.\ E}
  \textbf{\bibinfo{volume}{75}}, \bibinfo{pages}{011702}
  (\bibinfo{year}{2007}).

\bibitem[{\citenamefont{Fernsler et~al.}(2005)\citenamefont{Fernsler, Hough,
  Shao, Maclennan, Navailles, Madhusudana, Mondain-Monval, Boyer, Zasadzinski,
  Rego et~al.}}]{PNAS}
\bibinfo{author}{\bibfnamefont{J.}~\bibnamefont{Fernsler}},
  \bibinfo{author}{\bibfnamefont{L.}~\bibnamefont{Hough}},
  \bibinfo{author}{\bibfnamefont{R.-F.} \bibnamefont{Shao}},
  \bibinfo{author}{\bibfnamefont{J.~E.} \bibnamefont{Maclennan}},
  \bibinfo{author}{\bibfnamefont{L.}~\bibnamefont{Navailles}},
  \bibinfo{author}{\bibfnamefont{M.~B. N.~V.} \bibnamefont{Madhusudana}},
  \bibinfo{author}{\bibfnamefont{O.}~\bibnamefont{Mondain-Monval}},
  \bibinfo{author}{\bibfnamefont{C.}~\bibnamefont{Boyer}},
  \bibinfo{author}{\bibfnamefont{J.}~\bibnamefont{Zasadzinski}},
  \bibinfo{author}{\bibfnamefont{J.~A.} \bibnamefont{Rego}},
  \bibnamefont{et~al.}, \bibinfo{journal}{Proc. Nat. Acad. Sci. U.S.A.}
  \textbf{\bibinfo{volume}{102}}, \bibinfo{pages}{14191}
  (\bibinfo{year}{2005}).

\bibitem[{\citenamefont{Kamien and Lubensky}(1999)}]{Kamien}
\bibinfo{author}{\bibfnamefont{R.~D.} \bibnamefont{Kamien}} \bibnamefont{and}
  \bibinfo{author}{\bibfnamefont{T.~C.} \bibnamefont{Lubensky}},
  \bibinfo{journal}{Phys.\ Rev.\ Lett.} \textbf{\bibinfo{volume}{82}},
  \bibinfo{pages}{2892} (\bibinfo{year}{1999}).

\bibitem[{\citenamefont{Ogawa}(in Japanese)}]{OgawaPhD}
\bibinfo{author}{\bibfnamefont{H.}~\bibnamefont{Ogawa}}, Ph.D. thesis,
  \bibinfo{school}{Tohoku University, 2009} (\bibinfo{year}{in Japanese}).

\bibitem[{\citenamefont{Ogawa and Uchida}(2006)}]{OgawaTGB}
\bibinfo{author}{\bibfnamefont{H.}~\bibnamefont{Ogawa}} \bibnamefont{and}
  \bibinfo{author}{\bibfnamefont{N.}~\bibnamefont{Uchida}},
  \bibinfo{journal}{Phys.\ Rev.\ E} \textbf{\bibinfo{volume}{73}},
  \bibinfo{pages}{060701(R)} (\bibinfo{year}{2006}).

\bibitem[{\citenamefont{Kamien}(1997)}]{KamienDT}
\bibinfo{author}{\bibfnamefont{R.~D.} \bibnamefont{Kamien}},
  \bibinfo{journal}{J. Phys. II (France)} \textbf{\bibinfo{volume}{7}},
  \bibinfo{pages}{743} (\bibinfo{year}{1997}).

\bibitem[{\citenamefont{Dozov}(1995)}]{Dozov}
\bibinfo{author}{\bibfnamefont{I.}~\bibnamefont{Dozov}},
  \bibinfo{journal}{Phys. Rev. Lett.} \textbf{\bibinfo{volume}{74}},
  \bibinfo{pages}{4245} (\bibinfo{year}{1995}).

\bibitem[{\citenamefont{Koshelev}(1999)}]{Koshelev}
\bibinfo{author}{\bibfnamefont{A.~E.} \bibnamefont{Koshelev}},
  \bibinfo{journal}{Phys.\ Rev.\ Lett.} \textbf{\bibinfo{volume}{83}},
  \bibinfo{pages}{187} (\bibinfo{year}{1999}).

\end{thebibliography}

\end{document}